\documentclass[reprint,
superscriptaddress,
 amsmath,amssymb,
 aps,
showkeys
]{revtex4-1}

\makeatletter
\newcommand\emailx[1]{%
\move@AF%
\def\@affil{{\normalfont\,#1\strut}{}}%
}%

\usepackage{comment}
\usepackage{graphicx}
\usepackage{dcolumn}
\usepackage{bm}
\usepackage{hyperref}
\usepackage{physics}
\usepackage{mathtools}
\usepackage{subcaption}

\usepackage{graphicx,amssymb,amsmath,amsthm,amsfonts}
\usepackage[usenames]{color}
\usepackage{mathtools}

\begin{document}


\title{Gibbons\hskip0.03cm-\hskip-0.03cm{}Hawking
action for electrically charged black holes
in the canonical ensemble and Davies' thermodynamic theory of black
holes}



\author{Tiago V. Fernandes}
\email{tiago.vasques.fernandes@tecnico.ulisboa.pt}
\affiliation{Centro de Astrof\'isica e Gravita\c c\~ao 
- CENTRA, Departamento de F\'isica, Instituto Superior T\'{e}cnico 
- IST, Universidade de Lisboa - UL,\\
 Avenida Rovisco Pais 1, 1049-001 Lisboa, Portugal}
\author{Jos\'{e} P. S. Lemos}
\email{joselemos@ist.utl.pt}
\affiliation{Centro de Astrof\'isica e Gravita\c c\~ao 
- CENTRA, Departamento de F\'isica, Instituto Superior T\'{e}cnico 
- IST, Universidade de Lisboa - UL,\\ 
Avenida Rovisco Pais 1, 1049-001 Lisboa, Portugal}


\begin{abstract}

We establish the connection between the
Gibbons\hskip0.02cm-\hskip-0.02cm{}Hawking Euclidean path integral
approach applied to the canonical ensemble of a Reissner-Nordstr\"om
black hole and the thermodynamic theory of black holes of Davies. 
We build the ensemble, characterized by a reservoir at infinity at
temperature $T$ and electric charge $Q$, in $d$ dimensions. The
Euclidean path integral yields the action and partition function. In
zero loop, we uncover two solutions, one with horizon radius $r_{+1}$
the least massive, the other with $r_{+2}$, both meeting at a saddle point 
with radius $r_{+s}$ at temperature $T_s$.
We derive the thermodynamics, finding that the heat capacity diverges
at the turning point $T_s$ for each solution.
The free energy of the stable
solution is positive, so if the system is a black hole it makes a
transition to hot flat space with charge at infinity.
For a given $Q$ and
$T>T_s$, there is only hot space. An interpretation of the results
as energy wavelengths is attempted. For $d=4$, the thermodynamics from
the path integral applied to the canonical ensemble is precisely the
Davies thermodynamics theory of black holes, with $T_s$ being the
Davies point. We sketch the case $d=5$.

\end{abstract}


\maketitle

\section{\label{sec:Intro}Introduction}

The Euclidean path integral methods of statistical mechanics and
the theory of 
thermodynamics have been connected for a long time, a fact that
can be ascertained by consulting the great many works and books
dealing with this connection. It is not different with black holes,
the gravitational systems par excellence. The results obtained
through statistical methods applied to black hole spaces, in
juxtaposition to the results obtained through the analysis of quantum
fields around black holes, are related to black hole thermodynamic
theory and black hole entropy.

On one hand, applying the Euclidean path integral approach to statistical
mechanics of ensembles of spaces with a central black hole, Gibbons
and Hawking \cite{gibbhawk:1977} were able to identify the exact
temperature and entropy expressions of a black hole. In the Euclidean
path integral approach, the focus resides in constructing an ensemble
of a physical space. In order to build the ensemble, one needs a
microscopic description of the system a priori. For gravitation, an
accepted microscopic description is not yet known. Nevertheless, one
can choose the Euclidean path integral approach to quantum gravity to
compute the partition function of space.
In this approach, the statistical mechanics partition function $Z$ can
be written as a path integral of the exponential of the Euclidean
action $I[g,\phi]$, over Euclidean metrics $g$ and fields $\phi$ that
permeate the space, where the integral is restricted to metrics
$g$ that
are periodic in the imaginary time length, i.e., $Z=\int dg\, d\phi\,
{\rm e}^{-I[g,\phi]}$, with the fields
$\phi$ being periodic or anti-periodic
depending on their bosonic or fermionic character, respectively, 
and with the Planck's constant being set to unity.
To build the ensemble, one must fix some quantities for the fields at
some boundary. For instance, in the canonical ensemble, one fixes the 
temperature of the ensemble, which is given by the inverse of the 
imaginary time length at the boundary, with this boundary representing 
a heat reservoir.
However, when the partition function is computed in this manner,
one inherits the difficulties of the Euclidean path integral approach
to quantum gravity.
As an example, one needs to perform a map between the physical
Lorentzian spacetime and the Euclidean space, through a Wick
transformation of a time coordinate, but such map is not in general
well-defined or unique, covering only some sections of the Lorentzian
space.
For static and stationary spaces, the existence of a timelike Killing
vector field is used to specify the most convenient map to be
performed, as it is the case for static and stationary black hole
spaces. There are also obstacles in the convergence of the path
integral and, moreover, one can find difficulties in knowing what
classes of metrics one should integrate over.
To put these difficulties aside, one can consider the zero loop
approximation or saddle point approximation 
of the path integral, where only the paths that
minimize the Euclidean action are considered to
give the strongest
contributions to the path integral. The partition function should then
be given by $Z = {\rm e}^{-I_0}$, where $I_0$ is the classical Euclidean
action evaluated at one of these paths, and thus the procedure yields
a partition function in the semiclassical approximation since the
Planck's constant is hidden inside the exponential for $Z$.
For each ensemble, one considers that the logarithm of the partition
function yields the corresponding thermodynamic potential times the
inverse temperature. The thermodynamics of the system can then be
worked out through the derivatives of this thermodynamic potential.
The Gibbons\hskip0.02cm-\hskip-0.02cm{}Hawking results came from the
set of ideas explained above applied to a Schwarzschild black hole in
the canonical ensemble and to a Reissner-Nordstr\"om black in the
grand canonical ensemble, where here in addition to the temperature,
the electric potential is fixed at the boundary, and with the heat
reservoir placed at infinity. Although within the approach one could
recover the temperature and entropy of the corresponding black hole,
yet, in their respective ensembles, the heat capacity of the black
hole space is negative, meaning that the configuration is
thermodynamically unstable, and a thermodynamic treatment is thus not
valid.  It was later found that these configurations correspond to a
saddle point of the action \cite{Gross:1982}. Therefore, the zero loop
approximation cannot be applied to such configurations, although one
can still treat them as instantons. A perturbation of the instanton
yields a negative mode that makes the one-loop contribution of the
path integral imaginary, by an analytic continuation.
It was then noted that putting a black hole in a space with a negative
cosmological constant, which is
a system realized by the Schwarzschild-anti-de Sitter black hole in
the canonical ensemble, yielded one thermodynamically stable and
one thermodynamically unstable black hole solutions \cite{HawkingPage1983}. 
Notice that a space with negative cosmological constant 
acts in some extent as a box, in contrast with
the zero cosmological constant 
case.
It was also found that the negative mode of the Schwarzschild black
hole coming from spherically symmetric perturbations ceases to exist
if the heat reservoir, instead of being at infinity, sits at a radius
equal or smaller than the photon sphere radius~\cite{Allen:1984}.
York~\cite{York:1986} recognized that the construction of canonical or
grand canonical ensembles for a black hole space required a cavity
with a heat reservoir at finite radius. Through the path integral
approach and performing the zero loop approximation for a
Schwarzschild black hole inside a cavity, York found that there are
two stationary points, or black hole solutions, for the Euclidean
action $I_0$. From these two, the black hole with the least mass is unstable
and corresponds to the
Gibbons\hskip0.02cm-\hskip-0.02cm{}Hawking black hole in the limit of
infinite radius of the heat reservoir. The black hole with
the largest mass is stable, therefore the zero loop approximation is valid
for this stationary point. This result motivated a plethora of
developments in the Euclidean path integral approach
to quantum gravity and other related methods,
see, e.g.,
\cite{Whiting:1988,Martinez:1989,Braden:1990,Zaslavskii:1990,
Brown:1994,lwh1996,Lemos:1996,Peca:1999,chamblin:1999,
hawkreall:1999,Gregory:2001,carlipvaidya2003,Lundgren:2006,
Akbar:2010,hb2019,Andre:2020,
lizhangwang2020,Andre:2021,Miyashita:2021,jv:2023,Lemos:2023,
Fernandes:2023,lemoszaslavskii2024,ms2009}.

On the other hand, the idea that black holes are thermodynamic systems
was first hypothesized by Bekenstein~\cite{Bekenstein:1972}, by
equating the area of the event horizon of a black hole to its entropy,
which allowed for a formulation of a generalized second law of
thermodynamics.
This helped to indicate that the relations found for the mass, the
electric charge, and the angular momentum of a black hole, through the
Smarr formula~\cite{Smarr:1973} and its extension to the four laws of
black hole mechanics~\cite{Bardeen:1973}, were really
thermodynamic laws.
That black holes are thermodynamic objects was finally established by
Hawking~\cite{Hawking:1975}, with the discovery that black holes emit 
radiation, when quantum fields are present
in a black hole background. 
This radiation has a thermal spectrum at temperature
$T_{\rm H} = \frac{\kappa}{2\pi}$, the Hawking temperature in Planck
units, where $\kappa$ is the black hole surface gravity.
Admitting that the black hole is in equilibrium with the thermal
radiation, one concludes that the
whole system must be at the Hawking
temperature. Then, from the first law,
one obtains that black holes have entropy $S$ given by 
$S=\frac{A_+}{4}$, which is the Bekenstein-Hawking entropy, with $A_+$
being the surface area of the event horizon \cite{Hawking:1976}.
This thermodynamic account 
was put in a firm basis by the disclosure
that the appropriate vacuum state for the quantum field at the horizon
is described by the Hartle-Hawking vacuum
state~\cite{Hartle:1976,GPerry:1978}.
With these results and using the first law of black hole
thermodynamics, Davies~\cite{Davies:1977} was able to develop a
thermodynamic theory of the Kerr-Newman family of black holes, i.e.,
black hole solutions in general relativity with mass, electric charge,
and angular momentum, detailing it to Reissner-Nordstr\"om black
holes, with zero angular momentum, and to Kerr
black holes, with zero electric charge.
The thermodynamic theory for Reissner-Nordstr\"om black
holes was also established in \cite{hut1977}
almost concomitantly.
Within
the theory, it was found that the
Kerr-Newman black holes, and in particular, 
the Reissner-Nordstr\"om and the
Kerr black holes, have two branches, one with
 smaller horizon radius
and thermodynamically stable and the other with
larger horizon radius and
thermodynamically unstable. 
The Schwarzschild black hole would be reduced to the unstable branch
alone and could not count as a proper thermodynamic system.
Appropriately analyzed, one could detect an important point at
the conjunction of the two black hole solutions, which was interpreted
by Davies as a second order phase transition between the two black hole
branches, and by other authors as a turning point of stability, see
\cite{Davies:1978,sokomazur:1980,Davies:1989,kok1993}.
The first law of black hole thermodynamics has since been used in many
instances to understand black hole solutions, their
structure and stability, see, e.g.,
\cite{katz1993,parent,gpp,myung,maeda,la,hendi2014,cg:2020,
hajian2021,jiang2021,rodriguez,murk2023}.

Remarkably, there are instances where the Euclidean path integral
approach of gravitational systems with its associated ensemble
formulation,
and the thermodynamics of those systems with their
corresponding first law, get  entangled and share similarities.
Such similarities were found in the thermodynamics of hot thin shells
of various types of matter and different outer spaces
\cite{Martinez:1996,Lemos:2015a,Andre:2019,Reyes:2022,Fernandes:2022}
and in the quasiblack hole approach where the thermodynamic properties
of matter and spacetime on the verge of becoming a black hole are
ingeniously taken into account
\cite{Lemos:2009,Lemos:2010,Lemos:2020}.

It is our aim in this work to show that the
Gibbons\hskip0.02cm-\hskip-0.02cm{}Hawking Euclidean
path integral approach for the canonical ensemble of a
Reissner-Nordstr\"om black hole yields the Davies' thermodynamic
theory of the same black hole. The canonical ensemble of the pure
Reissner-Nordstr\"om black hole space is constructed for arbitrary
dimensions $d\geq4$ in such a manner
that the temperature and the electric charge of the reservoir are fixed
at infinity. We obtain the
partition function through the Euclidean path integral approach in the
zero loop approximation.
We find two solutions that minimize the action, with
both existing for an ensemble temperature $T$ less than
some critical or saddle value.
One solution
has a horizon radius below a given value, while
the other has its radius above that same given value.
For an ensemble temperature $T$ greater than
the critical or saddle value, there is only hot flat space solution
with electric charge at infinity.
We obtain the action for
the two black hole solutions. This action times the temperature is
the Helmholtz free energy of the system.  
A thermodynamic analysis of the solutions using the free energy shows
that the smaller solution is stable while the larger solution is
unstable, with the saddle point at the conjunction of both solutions
signalling the occurrence of a turning
point. We show that this saddle or critical point,
represents the generalization of the
Davies point to any dimension $d\geq4$.
The Helmholtz free energy permits also to compare
energetically the
stable black hole phase against the hot flat space phase
with electric charge at infinity, allowing to
determine which of the two phases is preferred.
It is possible to depict the main features
of the solutions in 
terms of
wavelengths of packets of thermal energy.
For $d=4$, studied in detail,
we recover the Davies' thermodynamic theory of black holes for
the first time from the relevant canonical ensemble. We also display
the case $d=5$ as a typical higher dimensional case.

This paper is organized as follows. In Sec.~\ref{sec:Canonical1}, we
construct the canonical ensemble through the Euclidean path integral
approach, giving the metric and the Maxwell tensor, analyze the
solutions of the ensemble, and compute their action in $d$
dimensions.  In Sec.~\ref{sec:thermo}, we study, in all
its details, the thermodynamics 
provided by the canonical ensemble in the zero loop
approximation in
$d$ dimensions.  In Sec.~\ref{sec:Daviesd4}, we recover Davies
thermodynamic results from the canonical ensemble by considering
$d=4$.  In Sec.~\ref{sec:Daviesd5} we consider $d=5$ to understand the
extension of these results.  In Sec.~\ref{sec:Concl}, we conclude.
We use units in which the constant of gravitation,
the speed of light, the Planck constant, and the Boltzmann constant
are set to one.

\vskip -1em

\section{\label{sec:Canonical1}
The canonical ensemble of a charged black hole in asymptotically flat
space through the Euclidean path integral approach}

\subsection{\label{sec:partition}Partition function given by the
Euclidean path integral approach}

The canonical ensemble of an electrically charged system in statistical
mechanics is a statistical ensemble composed by all possible
configurations of the system which are in thermal equilibrium with a
reservoir at fixed temperature $T$ and fixed electric charge $Q$, with
unspecified energy. The canonical ensemble allows one to obtain the
thermodynamic properties of the system in equilibrium with the
reservoir. The master quantity with all the thermodynamic
information in the canonical ensemble is the partition function $Z$
defined as $Z = \sum_i {\rm e}^{-\frac{E_i}{T}}$, where the sum of all
the possible states $i$ is done, $E_i$ is the energy of each state
$i$, and $T$ is the ensemble temperature.
In quantum mechanics, the partition function $Z$ can be expressed as
$Z = {\rm Tr}\,({\rm e}^{-\frac{H}{T}})$, where ${\rm Tr}\,({\rm
e}^{-\frac{H}{T}})$ is the trace of the quantum operator ${\rm
e}^{-\frac{H}{T}}$, $H$ being the Hamiltonian of the system expressed in
some Hilbert field space basis.
Using the Euclidean Feynman path integral action approach one finds
that for an imaginary time $\beta$ one has that for some field $\phi$,
${\rm Tr}({\rm e}^{-\beta H}) = \int d{\phi}\, {\rm e}^{-I[{\phi}]}$,
where $I$ is the Euclidean action of the system and the integration is
done for every possible periodic function ${\phi}$ in the bosonic
field case and for every possible antiperiodic
function ${\phi}$
in the fermionic case.
Thus, upon identifying $\beta$ with the inverse
temperature $T$ of the ensemble, $\beta=\frac1T$, the partition
function can be formally calculated through the path integral
expression $Z = \int d{\phi}\, {\rm e}^{-I[{\phi}]}$.  This
prescription works fine for 
matter fields $\phi$.
One can extend this approach to construct canonical
ensembles of self-gravitating systems, where the Euclidean metric $g$,
obtained from the Lorentzian metric by a Wick rotation to imaginary
time, plays the role of the field. Moreover, such an extension
provides a way to describe quantum gravity, i.e.,
it yields the Euclidean path
integral approach to quantum gravity.  The partition function can be
written then by the Euclidean path integral
\begin{align}
Z = \int dg \,{\rm e}^{-I[g]}\,,
\label{eq:partigionfunctionactiongrav}
\end{align}
where $I[g]$ is the action properly Euclideanized.

We now use the
Gibbons\hskip0.02cm-\hskip-0.02cm{}Hawking Euclidean path integral approach to
construct the canonical ensemble of an asymptotically flat spherically
symmetric electrically charged black hole space in arbitrary $d$
dimensions.  The system is in equilibrium with a heat reservoir at
infinity with temperature $T$ and electric charge $Q$.

\subsection{The Euclidean action for the canonical ensemble}

The Euclidean action of the system consisting of an
electrically charged black hole 
in asymptotically flat space in $d$ dimensions is
\begin{align}
    I =& - \frac{1}{16\pi}\int_M R \sqrt{g}\,d^d x 
- \frac{1}{8\pi} \int_{\partial M} (K-K_0)\sqrt{\gamma}\, d^{d-1}x\nonumber\\
&+ \frac{(d-3)}{4\Omega_{d-2}}\int_M F_{ab}F^{ab}\sqrt{g}\,d^dx\nonumber\\
&+ \frac{(d-3)}{\Omega_{d-2}}\int_{\partial M}F^{ab}A_{a}n_b 
    \sqrt{\gamma}\,d^{d-1}x\,,
    \label{eq:action1}
\end{align}
where $R$ is the Ricci scalar given by first and second 
derivatives of the Euclidean metric $g_{ab}$, 
$g$ is the determinant of $g_{ab}$,
$K$ is the trace of the 
extrinsic curvature $K_{ab}$ of the space boundary, 
$K_0$ is the trace of the extrinsic curvature 
of the space boundary
embedded in flat Euclidean space, 
$\gamma_{\alpha\beta}$
is the induced metric on the space boundary, 
$\gamma$ is the determinant of $\gamma_{\alpha\beta}$, 
$\Omega_{d-2} = \frac{2\pi^{\frac{d-1}{2}}}
{\Gamma\left(\frac{d-1}{2}\right)}$
 is the surface area of the unit $(d-2)$-sphere,
$F_{ab} = \partial_a A_b - \partial_b A_a$ is the Maxwell
tensor given by derivatives of the electromagnetic vector
potential $A_a$, 
and $n_b$ is the outward unit normal vector to the space boundary.
The 
Gibbons\hskip0.02cm-\hskip-0.02cm{}Hawking-York boundary term 
involving the extrinsic curvature
must be present in the action in order
to have a well-defined variational
principle with Dirichlet boundary conditions.
The boundary term depending on the Maxwell tensor 
must be present so that the canonical 
ensemble may be prescribed, 
see \cite{Braden:1990}.
This can be seen from the variation of the action, as one must 
get a boundary term with the variation of flux density and not the 
variation of the Maxwell field.  
This term gives the correct identification of the action with
fixed electric flux given by the 
integral of the Maxwell tensor on a $(d-2)$-surface, 
which has the meaning of electric charge.
In the other situation, the potential 
vector 
$A_a$ must be fixed
at the boundary in order to have a well-described system, 
which means the grand canonical ensemble should be prescribed
as was done in \cite{gibbhawk:1977}, see also \cite{Braden:1990}
and \cite{Fernandes:2023}.

\subsection{The zero loop approximation}

\subsubsection{Euclidean Reissner-Nordstr\"om line element and 
Maxwell field}

We now apply directly the zero loop approximation in the
Gibbons\hskip0.02cm-\hskip-0.02cm{}Hawking
sense \cite{gibbhawk:1977}, meaning that we evaluate 
the action in Eq.~\eqref{eq:action1} for a space that is a 
solution to the Euclidean Einstein-Maxwell equations. 
This solution for arbitrary $d$ dimensions
with $d\geq 4$,
is described by the $d$-dimensional
Reissner-Nordstr\"om line element
\begin{align}
    ds^2 =  \left(\frac{1}{2\pi\, t_{\rm H}}\right)^{\hskip -0.089 cm 2}
    \hskip -0.08cm f(r)\,
    d\tau^2 + \frac{dr^2}{f(r)} + r^2 d\Omega_{d-2}^2\,,
    \label{eq:ReissnerNordstrom}
\end{align}
also called 
Tangherlini
line 
element, where
the function $t_{\rm H}$, the Hawking function
or Hawking temperature function, is given by
\begin{align}
t_{\rm H} =
\frac{(d-3)\left(r_+^{d-3}  -
   \frac{\mu q^2}{r_+^{d-3}}\right)}
 {4\pi r_+^{d-2}}
\,,
    \label{eq:betahawking}
\end{align}
with $r_+$ being the horizon radius 
of the
black hole and $q$ its electric charge, the function
$f(r)$ is defined by
\begin{align}
    f(r) = \left(1 - \frac{r_+^{d-3}}{r^{d-3}} \right)
    \left(1 - \frac{\mu q^2}{r_+^{d-3}r^{d-3}} \right),
    \label{eq:f}
\end{align}
with
\begin{align}
\mu = \frac{8\pi}{(d-2)\Omega_{d-2}}\,,
\label{eq:mu}
\end{align}
and $d\Omega_{d-2}^2$ is the line element of the $(d-2)$-sphere 
with surface area $\Omega_{d-2} = \frac{2\pi^{\frac{d-1}{2}}}
{\Gamma\left(\frac{d-1}{2}\right)}$. The coordinate
range for the Euclidean time is $\tau \in \,]0,2\pi[$, the 
range for the radius coordinate is
$r\in \,]r_+,\infty[$, and the ranges of the 
angular coordinates are the usual ones.
The Maxwell electromagnetic potential field is described by 
\begin{align}
    A_\tau(y) = - \frac{iq}{2\pi (d-3)t_{\rm H} }\left(
        \frac{1}{r_+^{d-3}} - \frac{1}{r^{d-3}}
     \right)\,.
    \label{eq:maxwellfield}
\end{align}
We now discuss the considerations used to obtain the
precise forms of the line element and
the Maxwell field
given in Eqs.~\eqref{eq:betahawking}-\eqref{eq:maxwellfield}.
First for the line element, we choose a smooth metric,
i.e., the metric cannot have conical singularities or curvature
singularities.  In order to avoid a conical singularity at the horizon
and since we have chosen $2\pi$ periodicity in the Euclidean time, we
added the factor $\frac{1}{(2\pi t_{\rm H})^2} = 
\frac{1}{(\sqrt{f}\partial_r \sqrt{f})^{2}_{y=0}}$ 
to the usual $\tau \tau$
component of the Reissner-Nordstr\"om line element.  Secondly, for the
Maxwell field, we have chosen a gauge for $A_\mu$ such that only
$A_\tau$ is non-zero and where we assumed the nonexistence of magnetic
monopoles. Also, we have chosen the gauge such that
$A_\mu(r_+)=0$. This is to tie the Maxwell field measured by an Eulerian 
observer exactly to the
physical electric potential
given by $(d-3) \frac{2\pi i t_{\rm H}}{\sqrt{f}} A_\tau$, 
which should be bounded at the horizon.

The Reissner-Nordstr\"om line element characterized
by Eqs.~\eqref{eq:ReissnerNordstrom}-\eqref{eq:mu}
has several features. As we wrote it,
the main features are the two parameters,
namely, the horizon radius
$r_+$, and the electric
charge $q$.
There is an instance where the line element 
is characterized by one parameter alone,
instead of two, which is the case 
when the following equality holds
\begin{align}
r_{+e}= (\mu q^2)^{\frac{1}{2d-6}} \,,
\label{eq:extremalr}
\end{align}
in which case the horizon is called extremal.
From Eq.~\eqref{eq:extremalr} we see
that for a given electric charge $q$ 
the extremal horizon radius 
$r_{+e}$ has a precise value, or inverting
the equation we find that for a given horizon
radius $r_+$, there is an extremal electric
charge $q_e$ given by
$q_e=\sqrt{
\frac{r_+^{2d-6}}{\mu}}$.
Sometimes, it is of interest
to trade the horizon radius $r_+$ for
the space mass $m$ and the electric charge $q$ as
\begin{align}
    r_+ = \left(\mu m + \sqrt{\mu^2 m^2 - \mu q^2}
    \right)^{\frac{1}{d-3}}.
    \label{eq:r+asmandq}
\end{align}
This equation can be inverted to
give
$m = \frac{r_+^{d-3}}{2\mu} + \frac{q^2}{2r_+^{d-3}}$.
In terms of the mass, the extremal black hole
of Eq.~\eqref{eq:extremalr} obeys the relation
$\sqrt\mu\,m=q$, where here $q$ means the absolute
value of the electric charge.

\subsubsection{The ensemble and its solutions}

The canonical ensemble we are working with
has its boundary at infinity and 
is characterized by a reservoir
with a given temperature $T$ and with a given
electric charge $Q$. 
The inverse temperature at infinity, $\beta$, 
is determined by the Euclidean proper time at the 
boundary of the space, i.e., $\beta = 2\pi 
\left(\frac{\sqrt{f}}{2\pi t_{\rm H}}\right)_{r\rightarrow \infty}$. 
Using that $f(r\rightarrow \infty)=1$, we have that 
$\beta$ must be equal to the inverse
of the Hawking function $t_{\rm H}$.
Now, from the path integral formalism,
$\beta$ is the inverse temperature of the ensemble,
$\beta=\frac1T$, and so the ensemble temperature $T$
and the Hawking temperature function $t_{\rm H}(r_+,q)$
of Eq.~\eqref{eq:betahawking}
satisfy the relation
$
    T=t_{\rm H}(r_+,q)
$.
Notice that, since the period of the Euclidean
time $\tau$ is $2\pi$, 
the factor $(2\pi t_{\rm H})^{-2}$ was introduced 
on the time-time component of the metric in order to 
have regularity, therefore one often links the temperature 
function $t_{\rm H}$ 
to the avoidance of a conical singularity at the horizon 
if the Einstein equations are solved. 
In addition, in this canonical ensemble,
the electric flux $\int F^{ab}dS_{ab} = -i
\frac{\Omega_{d-2}}{2}Q$ or, equivalently,
the total electric charge, with the reservoir 
at spatial infinity, is fixed
to be $Q$ so that
the electric charge of the black hole $q$ obeys
$q=Q$.
In brief, the canonical ensemble we are considering
with fixed temperature $T$ and fixed electric charge $Q$
at infinity imposes the following
constraints to the possible black hole solutions,
\begin{align}
    T=t_{\rm H}(r_+,Q) \,,
 \label{eq:Hawkbeta}
\end{align}
\begin{align}
    Q=q \,.
 \label{eq:QqHawk}
 \end{align}
The latter equation tells us that the 
black holes that are solutions of
this ensemble must have their electric
charge $q$ equal to the ensemble electric charge $Q$.

Inverting Eqs.~\eqref{eq:Hawkbeta}
and \eqref{eq:QqHawk}
we see that the black hole solutions
that the ensemble may have
are of the generic form
\begin{align}
r_+=r_+(T,Q)\,,
 \label{eq:Hawkbetainverted}
\end{align}
\begin{align}
q=q(T,Q)\,,
 \label{eq:QqHawkinverted}
\end{align}
this later one having a direct solution 
$q=Q$.
Specifically, by rearranging Eq.~\eqref{eq:Hawkbeta} and
taking into account Eq.~\eqref{eq:QqHawk},
the black hole
solutions $r_+$, which are
formally represented in Eq.~\eqref{eq:Hawkbetainverted},
obey
\begin{align}
\left(\frac{d-3}{4\pi T}\right)(r_+^{2d-6} -
\mu Q^2) - r_+^{2d-5} = 0\,.
\label{eq:rpinT}
\end{align}
This equation,
Eq.~\eqref{eq:rpinT},
is not solvable analytically
for generic $d$.
However, one can perform an analysis of 
its solutions.
The function $t_{\rm H}(r_+,Q)$
in Eq.~\eqref{eq:betahawking}, see also Eq.~\eqref{eq:Hawkbeta}, 
has a maximum at 
\begin{align}
    r_{+s}= \left(\sqrt{(2d-5)\mu}\,Q\right)^\frac{1}{d-3}
    \label{r+davies}\,,
\end{align}
which is a saddle point of the 
action for the black hole, and
where from now onwards, $Q$
stands for the absolute value of the electric charge
$Q$ itself.
This saddle point of the 
action of the black hole has temperature 
\begin{align}
T_s=
\frac{(d-3)^2}
{2\pi (2d-5)
(\sqrt{(2d-5)\mu}\,Q)^{\frac{1}{d-3}}}
\label{Tdavies}\,.
\end{align}
Eliminating $Q$ in Eqs.~\eqref{r+davies} and
\eqref{Tdavies} one finds $r_{+s}$ in terms of
a given temperature $T$,
$r_{+s}=\frac{(d-3)^2}{2\pi (2d-5)T}$,
or inverting, for a given $r_+$,
one finds 
$T_s=\frac{(d-3)^2}{2\pi (2d-5)r_{+}}$.
In $d=4$, the temperature $T_s$ in Eq.~\eqref{Tdavies}
reduces to the Davies temperature,
and so, one can see Eq.~\eqref{Tdavies} as the generalization of 
the Davies temperature to $d$ dimensions.
By inspection of Eq.~\eqref{eq:rpinT},
for temperatures in the
interval $0 <T \leq T_s$,
there are two solutions,
the solution
$r_{+1}(T,Q)$ and the solution $r_{+2}(T,Q)$, while 
for $T > T_s$
there are no black hole solutions.
The solution
$r_{+1}(T,Q)$ exists in the interval 
$r_{+e} <r_{+1}(T,Q)\leq r_{+s}$,
so we can summarize for the solution 1
\begin{equation}
\begin{aligned}
&r_{+1}=r_{+1}(T,Q),\quad\quad 0 <T \leq T_s,
\\
& q_1=Q,\quad\quad\quad\quad\quad\quad\,
r_{+e} <r_{+1}(T,Q)\leq r_{+s},
\end{aligned}
\label{solutionr+1}
\end{equation}
where 
$r_{+e}$ is the
radius of the extremal black hole
given by 
$r_{+e}=r_{+1}(T\rightarrow 0,Q) = (\mu Q^2)^{\frac{1}{2d-6}}$,
see Eq.~\eqref{eq:extremalr},
and 
$ r_{+s}=r_{+1}(T_s,Q)$ is given in Eq.~\eqref{r+davies}.
This solution,
$r_{+1}(T,Q)$, is an increasing monotonic
function of $T$.
The solution $r_{+2}(T,Q)$
exists in the interval 
$r_{+s}< r_{+2}(T,Q)<\infty$,
so we can summarize for the solution 2
\begin{equation}
\begin{aligned}
&r_{+2}=r_{+2}(T,Q),\quad\quad 0 <T \leq T_s,
\\
& q_2=Q,\quad\quad\quad\quad\quad\quad\,
r_{+s}< r_{+2}(T,Q)<\infty,
\end{aligned}
\label{solutionr+2}
\end{equation}
where 
$r_{+s}=r_{+2}(T_s,Q)$, i.e., the solution 2
is bounded 
from below, and
is unbounded from above, since 
at $T \rightarrow 0$, the 
solution tends to infinity. This solution,
$r_{+2}(T,Q)$, is a decreasing monotonic
function of $T$. When the ensemble is only characterized
by the temperature $T$, with $Q$ vanishing, $Q=0$,
only the black hole solution
$r_{+2}$ survives which corresponds to the
Gibbons\hskip0.02cm-\hskip-0.02cm{}Hawking black hole 
solution.
For $T>T_s$, there are no black hole solutions and one
is left with hot flat space with electric charged
$Q$
dispersed at infinity, and so the solution of the ensemble
at this temperature range can be summarized as
\begin{equation}
\begin{aligned}
&{\rm charged\; hot\;flat\;space},\quad\quad T_s<T<\infty,
\\
& Q \;{\rm dispersed \;at}\; r=\infty,\quad\quad
0\leq r<\infty.
\end{aligned}
\label{solutionhot}
\end{equation}
Thus, the three solutions of the ensemble are displayed
in Eqs.~\eqref{solutionr+1}-\eqref{solutionhot}.

\subsubsection{Action of the Reissner-Nordstr\"om black hole
space and partition function}

We now evaluate the action
given in Eq.~\eqref{eq:action1} for the metric
in Eq.~\eqref{eq:ReissnerNordstrom} and the Maxwell field in 
Eq.~\eqref{eq:maxwellfield}, with the
black hole solutions of the ensemble 
given by solving Eq.~\eqref{eq:rpinT},
i.e., those formally shown in  Eq.~\eqref{solutionr+1}
and Eq.~\eqref{solutionr+2}.
We sketch the calculations.

First, we consider the term depending on the Ricci scalar $R$.
The Ricci scalar for the metric in 
Eq.~\eqref{eq:ReissnerNordstrom} is given by
$-\frac{\sqrt{g}}{16\pi}R = 
    \frac{1}{8\pi}\left( \frac{r^{d-2}f'}{4\pi t_{\rm H}}\right)'
    - \frac{1}{4\pi t_{\rm H}}
    \frac{(d-3)q^2}{\Omega_{d-2}r^{d-2}}$,
where a dash means derivative with respect to $r$.
Second, with respect to the
Gibbons\hskip0.02cm-\hskip-0.02cm{}Hawking-York boundary term given
by $-\eval{\frac{\sqrt{\gamma}}{8\pi}(K-K_0)}$, one can evaluate it at
infinity to find
$-\eval{\frac{\sqrt{\gamma}}{8\pi}(K-K_0)}_{r\rightarrow\infty} =
\eval{\left[\frac{\sqrt{f} r^{d-3}} {2\pi t_{\rm H} \Omega_{d-2}
\mu}\left(1 - \sqrt{f}\right)\right]}_{r\rightarrow \infty} -
\frac{1}{8\pi}\eval{r^{d-2} \frac{
f'}{4\pi t_{\rm H}}}_{r\rightarrow\infty}$, where it was used that the
extrinsic curvature of a constant $r$ hypersurface is $\mathbf{K}
=\frac{\sqrt{f} f'}{2(2\pi t_{\rm H})^2}d\tau + r \sqrt{f} d\Omega_{d-2}^2$
and that $\mathbf{K}_0=r d\Omega_{d-2}^2$ is the extrinsic curvature
of the hypersurface embedded in flat space.
Third, with respect to the Maxwell boundary term, it is useful to
rewrite it in the action
Eq.~\eqref{eq:action1}. Using the regularity condition $A_{\tau}(r_+)
= 0$, the divergence theorem and that $\nabla_b(F^{ab}A_a) = -
\frac{1}{2} F_{ab}F^{ab} + A_a\nabla_bF^{ab} $, one transforms the
boundary Maxwell term into a bulk term, assuming the term inside the
divergence is bounded, obtaining that the Maxwell part of the action
is $- \frac{(d-3)}{4\Omega_{d-2}}\int_M F_{ab}F^{ab}\sqrt{g}d^dx +
\frac{(d-3)}{\Omega_{d-2}}\int_M A_a\nabla_b F^{ab} \sqrt{g}d^dx$.
Now, $-\frac{(d-3)\sqrt{g}}{4\Omega_{d-2}}F_{ab}F^{ab} = \frac{(d-3)}
{4\pi t_{\rm H} \Omega_{d-2}} \frac{q^2}{r^{d-2}}$,  where $F_{y\tau} =
A'_\tau$ was used, and $\frac{(d-3)\sqrt{g}}{\Omega_{d-2}}\nabla_b
F^{ab}A_{a} = 0$, since this is exactly a component of the Maxwell
equations.
Performing the limit 
$\eval{\left[\sqrt{f} r^{d-3}\left(1 -
\sqrt{f}\right)\right]}_{r\rightarrow \infty} = 
\frac{r_+^{d-3}}{2} + \frac{\mu q^2}{2r_+^{d-3}}$, 
one can proceed with the integrations to obtain, from
Eq.~\eqref{eq:action1}, the full zero loop action $I_0$, which depends
only on the ensemble parameters, the temperature
$T$ and the electric charge $Q$, namely,
\begin{align}
I_0[T,Q]
=
\frac{1}{\mu T}
    \left(\frac{r_+^{d-3}}{2} + \frac{\mu Q^2}{2r_+^{d-3}}\right) 
    - \frac{\Omega_{d-2}}{4}r_+^{d-2},
    \label{eq:actioninfinity}
\end{align}
where, we recall that $T=t_{\rm H}(r_+,Q)$ has two black hole solutions for
$T\geq T_s$, $r_{+1}(T,Q)$ and $r_{+2}(T,Q)$, each of which gives an
expression in terms of $T$ and $Q$ to Eq.~\eqref{eq:actioninfinity}.
Explicitly, the actions for each solution are of the form
$I_0(T,Q,r_{+1}(T,Q))$ and $I_0(T,Q,r_{+2}(T,Q))$.  There is a
third solution that must be considered, corresponding to the case of
having no black hole solutions.  This case is described by hot flat
space with fixed temperature of the reservoir at infinity and with
fixed electric charge residing at the reservoir at infinity, in order
to satisfy the Gauss constraint of the electromagnetic field without
contributing to the energy content of the space. This hot flat space
in this zero loop approximation is simply classical flat space at some
temperature $T$ with no matter fields present.  The zero loop action
for classical hot flat space with electric charge at infinity is then
zero, i.e., $I_0[T,Q]=0$. The partition function $Z$ in the zero loop
approximation for the canonical ensemble is then
\begin{align}
Z={\rm e}^{-I_0[T,Q]}\,,
    \label{eq:Zactionzeroloop}
\end{align}
with $I_0[T,Q]$ given in Eq.~\eqref{eq:actioninfinity}.

The partition function given in Eq.~\eqref{eq:Zactionzeroloop}
together with the action of Eq.~\eqref{eq:actioninfinity}
are valid in $d$ dimensions. 
In four dimensions, $d=4$, they 
will give origin to Davies results \cite{Davies:1977},
see also \cite{hut1977}. 
This means that
Davies' thermodynamic theory
of black holes,
in this case of electrically charged black holes, can be
seen within the canonical ensemble formalism. Here we
generalize the results to arbitrary $d$ dimensions, 
$d=4$ being a particular
case.

\section{Thermodynamics \label{sec:thermo}}

\subsection{Free energy,
entropy, electric potential, and  
thermal energy
\label{sec:thermoquantities}}

We have used the
Gibbons\hskip0.02cm-\hskip-0.02cm{}Hawking Euclidean path 
integral approach
to construct the canonical ensemble
of an asymptotically flat spherically symmetric
electrically charged 
black hole space 
in arbitrary $d$ dimensions.
With the system being in
equilibrium with a heat reservoir at
infinity with temperature $T$ and electric charge $Q$,
the thermodynamics of the system can now be obtained by
considering that the partition function of the canonical ensemble is
related to the Helmholtz free energy $F$ through $Z =
e^{-\frac{F}{T}}$, i.e., $F=-T\ln Z$.
From the zero loop approximation, 
Eq.~\eqref{eq:Zactionzeroloop},
this means $F=TI_0$.
With $I_0$ given in Eq.~\eqref{eq:actioninfinity}
one finds that the free energy is
$
F = \frac{1}{\mu}\left(\frac{r_+^{d-3}}{2} 
+ \frac{\mu Q^2}{2r_+^{d-3}}\right)
-\frac{\Omega_{d-2} r_+^{d-2}}{4}T$. 
Substituting $T$ for $t_{\rm H}$, see Eqs.~\eqref{eq:betahawking}
and \eqref{eq:Hawkbeta}, we obtain for
the free energy the expression
\begin{align}
F(T,Q)
\hskip-0.05cm
=
\hskip-0.05cm
\frac{1}{\mu (d-2)}\hskip-0.1cm
\left(
\hskip-0.05cm
\frac{r_{+}^{d-3}}{2} 
+ (2d-5)\frac{\mu Q^2}{2r_{+}^{d-3}}
\hskip-0.05cm
\right)\hskip-0.05cm,
\label{eq:freeenergyinfinity}
\end{align}
where $r_+$ should be envisaged
as  $r_+=r_+(T,Q)$, since it is one of
the solutions 
$r_{+1}(T,Q)$ or
$r_{+2}(T,Q)$, given in 
Eq.~\eqref{solutionr+1} or
Eq.~\eqref{solutionr+2}, respectively.
Thus, the Helmholtz free energy $F$
for each solution is a function only of $T$ and $Q$,
namely, 
$F(T,Q,r_{+1}(T,Q))$ and $F(T,Q,r_{+2}(T,Q))$.

With the free energy $F$ given by Eq.~\eqref{eq:freeenergyinfinity},
we can obtain the thermodynamic quantities 
through its differential,
$dF = -S dT + \phi dQ$. The first component of the differential 
yields the entropy 
\begin{align}
S = \frac14 A_+\,,
\end{align}
where $A_+=\Omega_{d-2} r_+^{d-2}$ is the
area of the horizon,
and so $S$ is the Bekenstein-Hawking entropy, valid
for the two solutions 
$r_{+1}(T,Q)$ or
$r_{+2}(T,Q)$.
The second component of the differential yields the thermodynamic
electric potential 
\begin{align}
\phi = \frac{Q}{r_+^{d-3}}\,,
\end{align}
i.e., the Coulombic electric potential,
with  $r_+$ being $r_{+1}(T,Q)$ or
$r_{+2}(T,Q)$.
The thermodynamic energy, given by $E = F + TS$,
has the form 
$E = \frac{r_+^{d-3}}{2\mu} + \frac{Q^2}{2r_+^{d-3}}$,
and since  $r_+$ is $r_{+1}(T,Q)$ or
$r_{+2}(T,Q)$, there are two solutions for $E$.
This can be connected to the space mass $m$
given by
$m = \frac{r_+^{d-3}}{2\mu} + \frac{Q^2}{2r_+^{d-3}}$,
see Eq.~\eqref{eq:r+asmandq},
so
that here the thermodynamic energy and the black hole
mass
are equal, i.e., they obey the relation
\begin{align}
E = m\,.
\label{eq:energyinfinity}
\end{align}
Thus, we can write the free energy
given in Eq.~\eqref{eq:freeenergyinfinity} as
\begin{align}
F=m-TS\,.
\label{eq:freeenergyinfinityagain}
\end{align}

\subsection{Smarr formula and the first law of black hole mechanics}

The energy in the
form
$E = \frac{r_+^{d-3}}{2\mu} + \frac{Q^2}{2r_+^{d-3}}$
can also be rewritten
in terms 
of the entropy and electric charge as
$E = \frac{1}{2\mu}\left(\frac{4S}{\Omega_{d-2}}
\right)^{\frac{d-3}{d-2}} 
+ \frac{Q^2}{2}\left(\frac{4 S}{\Omega_{d-2}}
\right)^{\frac{3-d}{d-2}}$.
This energy function has the scaling property
$\nu^{\frac{d-3}{d-2}} E = E(\nu S, \nu^{\frac{d-3}{2(d-2)}} Q)$,
for some $\nu$, and so
through the Euler relation theorem,
we have $E= \frac{d-3}{d-2} TS + \phi Q$.
Together with Eq.~\eqref{eq:energyinfinity}, i.e.
$E=m$, we obtain 
\begin{align}
m = \frac{d-3}{d-2} TS + \phi Q\,,
\label{smarrd}
\end{align}
which is the Smarr formula
for an electrically charged black hole in $d$ dimensions.
The Smarr formula is valid for the two solutions, $r_{+1}(T,Q)$ or
$r_{+2}(T,Q)$.

One can also verify that the first law
of thermodynamics,
\begin{align}
dm =  TdS + \phi dQ\,,
\label{eq:firstlawofbhm}
\end{align}
holds. It holds for the two solutions, $r_{+1}(T,Q)$ or
$r_{+2}(T,Q)$.
But, Eq.~\eqref{eq:firstlawofbhm} is also
the first law of black hole mechanics
since it involves pure black hole quantities.
This shows that
the thermodynamics that follow from the
electrically charged
canonical ensemble statistical mechanics
is equivalent to the thermodynamics that
follows from the first law of black hole mechanics.
The first law of black hole mechanics was the starting
point of Davies' analysis, while here it is a result of 
the  statistical mechanics formalism.

\subsection{Heat capacity and thermodynamic stability}

The thermodynamic stability of the system is given by the 
condition that the heat capacity
at constant electric charge must be positive, ensuring that 
the respective solution is stable. The heat capacity
at constant electric charge is
defined by 
$C_Q =\left( \frac{\partial E}{\partial T} \right)_Q$.
Since $E = m=\frac{r_+^{d-3}}{2\mu} + \frac{Q^2}{2r_+^{d-3}}$ and
$r_+=r_+(T,Q)$, we have
\begin{align}
C_Q& = \frac{(d-2)\Omega_{d-2} r_+^{d-2}(r_+^{2d-6}-\mu Q^2)}
{4\left((2d-5)\mu Q^2 - r_{+}^{2d-6} \right)}\nonumber\\
&=
\hskip-0.1cm
\frac{m S^3  T}
{\frac{(d
\hskip-0.02cm
-
\hskip-0.03cm
3)\Omega_{\hskip-0.03cm d-2}^3}{4^5 \pi^2}
\hskip-0.15cm
\left[
\frac{
(\hskip-0.02cm
3d-8
\hskip-0.02cm
)
\mu^2 Q^4}
{\left(\frac{4S}{\Omega_{d-2}}\right)^{\hskip-0.06cm\frac{d-4}{d-2}}} 
\hskip-0.05cm
+
\hskip-0.05cm
(d
\hskip-0.02cm
-
\hskip-0.02cm
4)
\hskip-0.06cm
\left(\frac{4 S}{\Omega_{d-2}}\right)^{\frac{3d-8}{d-2}}
\right]
\hskip-0.16cm
-
\hskip-0.11cm
T^2
\hskip-0.06cm
S^3}\hskip-0.02cm,
\label{heatcapacityraw}
\end{align}
where in the
second equality we wrote
the heat capacity in terms of the thermodynamic 
variables $m$, $S$, $T$, and $Q$.
We must note however that 
the heat capacity must be understood as a function of $\,T$ and 
$Q$, as these are the quantities controlled in the ensemble. 
This means that $r_+$ must be understood as either $r_{+1}(T,Q)$ 
or $r_{+2}(T,Q)$, as well as $m$ and $S$ must be understood as
$m=m(T,Q)$ and $S=S(T,Q)$.
The 
thermodynamic stability is satisfied if the heat capacity is 
positive. According to Eq.~\eqref{heatcapacityraw}, the ensemble is 
stable in the range $r_{+e}\leq r_+ < r_{+s}$, where 
$r_{+e}= (\mu Q^2)^{\frac{1}{2d-6}}$ and 
$r_{+s}= \left(\sqrt{(2d-5)\mu}\,Q\right)^\frac{1}{d-3}$.
This is precisely the range 
of the solution $r_{+1}$. Therefore, one has
\begin{align}
{\rm stability \,\, if}\; C_Q\geq0,\quad{\rm i.e.,} \quad
r_{+} = {r_{+1}}\,.
\label{eq:stability}
\end{align}
In opposition, the ensemble is unstable in the range 
$r_{+s}< r_+ < \infty$, which is the range of the solution $r_{+2}$. 
Hence, one has
\begin{align}
\hskip -0.3cm
{\rm instability \,\, if}\; C_Q<0,\quad{\rm i.e.,} \quad
r_+ = r_{+2}\,.
\label{eq:instability}
\end{align}
So, from  Eqs.~\eqref{eq:stability} and
\eqref{eq:instability}
one has that the solution $r_{+1}$ is stable
whereas the solution $r_{+2}$ is unstable, see
Eqs.~\eqref{solutionr+1} and \eqref{solutionr+2}. It must be noted
also that $r_{+1}$ is an increasing monotonic function of $T$, so that
the energy of the system increases when the temperature increases,
as it is expected from a stable system. The opposite happens to the
solution $r_{+2}$, since it is a decreasing monotonic function of $T$
and so the energy of the black hole decreases when the temperature
increases. 
From Eq.~\eqref{eq:stability}, we also find that the radius
${r_{+s}}$ acts as the generalization of the Davies point for higher
dimensions. Indeed, for ${r_{+s}}$ fixed, for steady addition of
electric charge $Q$, one finds that the solution passes from an
$r_{+2}$ solution to an $r_{+1}$ solution, and eventually at the
transition, a negative $C_Q$ turns into a positive $C_Q$.
In thermodynamics, this could signal a phase transition of second
order, since the free energy $F$ and its first derivatives are
continuous, but second derivatives are discontinuous. However, this is
not the case here, one is instead in the presence of a turning point
which determines the relative scale of $r_+$ and $Q$
at which a black hole can
be in stable or metastable equilibrium
when in thermal
contact with a heat reservoir that holds $T$ and $Q$
fixed at infinity.
Indeed, in the canonical ensemble, the parameters that one can
control are $T$ and $Q$.  Maintaining $Q$ fixed, and for a given
sufficiently low $T$, there are two solutions, the stable solution
$r_{+1}(T,Q)$ and the unstable solution $r_{+2}(T,Q)$. One could try 
to start with the stable solution at low $T$ and 
devise a change of parameters $T$ and $Q$ such that $r_+$ 
was kept fixed. Eventually, one is able to reach the turning point 
and beyond it, the character of the solutions changes, i.e., the 
unstable solution $r_{+2}$ would have a fixed $r_+$, 
while the stable solution $r_{+1}$ still exists 
and would suffer a change in $r_+$. 
But any thermal perturbations would make the unstable solution $r_{+2}$
to run away from equilibrium, thus the unstable solution $r_{+2}$ is 
impossible to be maintained. And so, even for this specific change of 
parameters, with temperature up to $T_s$,
one is always in the presence of the stable solution $r_{+2}(T,Q)$, 
this being the existing solution of the ensemble at $T_s$, 
and so one should not classify 
this point as a point of phase transition.

Bear in mind that the thermodynamic quantities,
the first law of thermodynamics, and the Smarr formula as an
integrated first law of thermodynamics, are strictly only
valid for the stable black hole
solution $r_{+1}$, since the solution $r_{+2}$ is unstable
and does not allow a proper thermodynamic treatment.
Note also, that in the limit of
zero electric charge, $Q=0$, there is only the $r_{+2}$ solution 
corresponding to the
Gibbons\hskip0.02cm-\hskip-0.02cm{}Hawking black hole solution 
which is unstable. 
Indeed, the heat capacity in the zero electric charge case is $C =
-\frac{(d-2)\Omega_{d-2} r_+^{d-2}} {4}$,
thus negative for all $r_+(T,Q)$.

\subsection{Favorable phases}

In a thermodynamic system, if different thermodynamic phases can take
place, it is of interest to know which are the favored phases for a
given set  of parameters.  For temperature $T$ and electric charge
$Q$ fixed by the reservoir, a thermodynamic system tends to be in a
state in which its Helmholtz free energy $F$ has the lowest value.  If
a system is in a stable state but with a higher free energy $F$ than
another stable state, it is probable that the system undergoes 
a transition, to the state with the lowest free
energy. Returning to the path integral calculation and the
corresponding partition function, we see that if there are two stable
configurations, i.e., two states that minimize the action, then the
largest contribution to the partition function is given by the
configuration with the lowest action or, in thermodynamic language, with the
lowest free energy. In order to analyze these phase transitions, one 
must obtain the critical regions where the free energy is the same 
for both configurations. Generally, at these transition points,
the free energy's derivatives 
are different, signaling first order phase transitions.

In the case of a cavity within a heat reservoir at infinity
kept at $T$ and $Q$ constants,
we have seen that within our context there are three solutions.
One is the stable black hole $r_{+1}$, Eq.~\eqref{solutionr+1}, which
counts as a thermodynamic phase and exists for $T\leq T_s$. The other
is the unstable black hole $r_{+2}$, Eq.~\eqref{solutionr+2}, which
also exists for $T\leq T_s$, but does not count as a thermodynamic
phase since it is unstable. The other is hot flat space with electric
charge at infinity
that exists for $T> T_s$, Eq.~\eqref{solutionhot}.  We
have taken this phase, where there are no black holes, to be hot flat
space with electric charge dispersed at infinity, because it seems
the most natural solution, as electric charge of the same sign repels,
and eventually disperse to infinity.

Thus, there are two possible phases, namely, the black hole $r_{+1}$
phase and hot flat space with electric charge at infinity.  For
$T>T_s$, only hot flat space with electric charge at infinity exists,
as we have seen.  But for $T<T_s$, both $r_{+1}$ and hot flat space
with electric charge at infinity can exist. The one that is going to
dominate for $T\leq T_s$ is the one that has the lowest free energy.  Now,
the free energy of hot flat space with electric charge at infinity is
zero,
\begin{align}
F_{\rm hfs}=0\,.
\label{eq:Fhfs}
\end{align}
The free energy of the $r_{+1}$ black hole is
always positive, $F(T,Q)= F(T,Q,r_{+1}(T,Q))>0$. This can be seen
from the on-shell expression
Eq.~\eqref{eq:freeenergyinfinity} which for the $r_{+1}$ solution reads
\begin{align}
F(T,Q,r_{+1}(T,Q))
\hskip-0.05cm
=
\hskip-0.05cm
\frac{1}{\mu (d-2)}\hskip-0.1cm
\left(
\hskip-0.05cm
\frac{r_{+1}^{d-3}}{2} 
+ (2d-5)\frac{\mu Q^2}{2r_{+1}^{d-3}}
\hskip-0.05cm
\right)\hskip-0.05cm.
\label{eq:Fr+1}
\end{align}
We find that
Eq.~\eqref{eq:Fr+1}
has strictly positive 
terms. Thus, since
\begin{align}
F_{\rm
hfs}<F(T,Q,r_{+1}(T,Q))\,,
\label{eq:comparisonFs}
\end{align}
hot flat space with electric charge at
infinity is the favored phase for $T\leq T_s$.
If
the system finds itself in the black hole phase, it will
make a transition to hot flat space with electric charge
at infinity
since it has lower free energy.
We note however
that the free energy of these two phases never intersects
and so we cannot call
this a first order phase transition. An analog
to this transition is the one between supercooled
water and ice.
Moreover, hot flat space 
it is the only phase 
for $T>T_s$.

\subsection{Interpretation}

We have deduced these thermodynamic results starting from the path
integral approach and then derived the thermodynamics from it. The
action that has entered into the path integral is the classical action
yielding thus a zero loop approximation.
Although in this order of
approximation there is no mention to matter fields, which would enter
in a first loop approximation, one can try to interpret some of the
results found in zero order, in terms of wavelengths of packets of
thermal energy inside the cavity of a heat reservoir at infinity.
This is because there is a given temperature $T$ within the system,
and we know that at a quantum level, for a given $T$, there is an
associated thermal wavelength $\lambda$, which is
$\lambda=\frac{(d-3)^2}{2\pi (2d-5)T}$.
We note that the interpretation of the results in terms of matter 
fields is useful as we will
see now,
even if it is beyond the formalism we use.

Let us start by interpreting the existence and nonexistence of the two
black hole solutions $r_{+1}$ and $r_{+2}$.
For small enough temperature $T$, and so large thermal wavelength
$\lambda$, there are two solutions for $r_+$.
The $r_+$ of the small solution is sufficiently small so that it is
smaller than $\lambda$, and so energy packets with typical wavelength
$\lambda$ are trapped in the black hole geometry and do not escape,
making the black hole a possible solution and a stable one.
The $r_+$ of the large solution is sufficiently large so that it is of
the order of $\lambda$, with $r_+$ being a bit larger,
and so energy packets with
typical wavelength $\lambda$ can escape, and backreact to turn the
black hole unstable.  Indeed, this case with $r_+$ of the order
$\frac1T$ and so of the order of $\lambda$, corresponds to the 
black hole with the
Gibbons\hskip0.02cm-\hskip-0.02cm{}Hawking
black hole solution properties.
Now, for larger reservoir temperature $T$, the
thermal wavelength $\lambda$ gets smaller. The $r_+$ of the small
solution is still small but now $r_+$ is barely smaller than
$\lambda$. The $r_+$ of the large solution
is now smaller than when the temperature was small,
with
$r_+$ being barely larger than $\lambda$. This latter solution
is still the one  with the
Gibbons\hskip0.02cm-\hskip-0.02cm{}Hawking 
black hole solution properties.
At a saddle or critical temperature $T_s$ the two solutions meet.
For even higher reservoir temperature
$T$, and so lower thermal wavelength
$\lambda$, there is no way to make a
black hole, the wavelength $\lambda$ is low enough that it
disperses without being able to aggregate the energy and the electric
charge in a black hole state.  In this case the electric charge
disperses to infinity, yielding hot flat space for the whole space
with the electric
charge at infinity, and so vanishing electric charge density.

We could try to interpret the favorable phases 
in terms of
wavelengths of packets of thermal energy inside the cavity, but we
have not found any direct way to see
how these packets yield that hot flat space with electric charge
at infinity has always, for all parameters,
a free energy lower than the small black hole free energy.
However, it is clear what happens when one looks at the
free energy expressions.
Looking at the
original expression for the free energy of the
stable black hole, i.e.,
$
F = \frac{1}{\mu}\left(\frac{r_+^{d-3}}{2} 
+ \frac{\mu Q^2}{2r_+^{d-3}}\right)
-\frac{\Omega_{d-2} r_+^{d-2}}{4}T$,
one sees that the entropy term which is negative has a small
contribution because $r_+$ is small, and there is the electric
charge term which goes as $\frac{Q^2}{2r_+^{d-3}}$ which
gives a large positive contribution, since $r_+$ is small,
all contributing for $F$ never being zero for any set of
parameters $T$ and $Q$.

To better understand all the issues
that we have worked out so far
and to make progress, one has to pick up
definite dimensions. We now specify our generic $d$-dimensional
results to the dimensions $d=4$ and $d=5$.  We
will do a thorough analysis for the dimension
$d=4$, and comment on the
dimension
$d=5$.

\section{The case $d=4$: 
Davies' thermodynamic theory of black holes and
Davies point from the canonical ensemble\label{sec:Daviesd4}}

\subsection{Solutions and action in $d=4$}

The dimension $d=4$ is specially interesting since it 
gives the results of Davies' thermodynamic theory of black holes
\cite{Davies:1977}, see also \cite{hut1977}. 

One must 
start from the canonical ensemble characterized by a heat reservoir
at infinity with
temperature $T$ and electric charge $Q$
 in $d=4$. 
The black hole solutions $r_+$
of the ensemble are taken from solving Eq.~\eqref{eq:Hawkbeta}
together with Eq.~\eqref{eq:betahawking},
which in $d=4$ they yield
\begin{align}
T=t_{\rm H}(r_+,Q),\quad\quad
t_{\rm H}(r_+,Q)=\frac{r_+-\frac{Q^2}{ r_+}}
{4\pi r_+^2}\,,
\label{eq:Hawkbeta0d=4}
\end{align}
where again
$T$ is the temperature kept fixed at the reservoir
at infinity and 
$t_{\rm H}(r_+,Q)$ is
the original Hawking function in $d=4$.
When the electric charge of the reservoir at infinity
is zero, $Q=0$, then 
$t_{\rm H}(r_+,Q)=\frac{1}
{4\pi r_+}$, which is the Hawking temperature of a
Schwarzschild black hole.
The electric charge $Q$ is 
the electric charge kept fixed at the reservoir
at infinity, and the black hole electric charge
$q$ 
must match it to have a consistent solution,
$q=Q$, see Eq.~\eqref{eq:QqHawk}.

\begin{figure}[h]
    \centering
    \includegraphics[width=\linewidth]{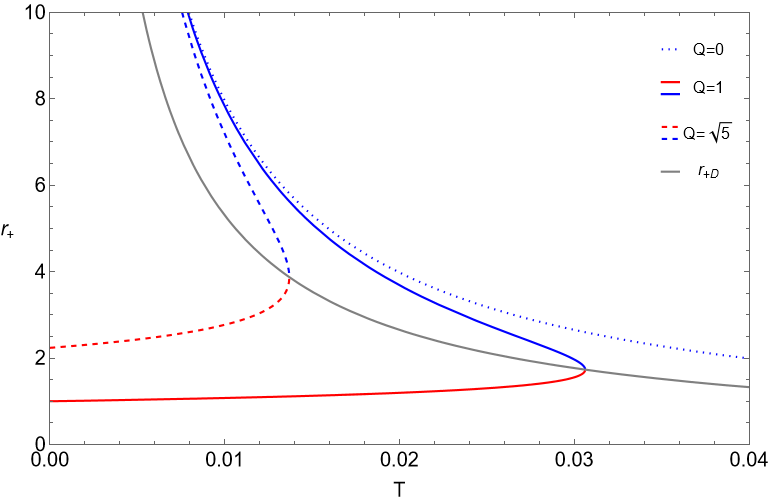}
    \caption{Plot of the two solutions $r_{+1}(T,Q)$, in red, and
    $r_{+2}(T,Q)$, in blue, of the charged black hole in the canonical
    ensemble for infinite cavity radius, as a function of $T$, for
    three values of the charge, $Q = 0$ in a dotted line, $Q = 1$ in
    filled lines, and $Q = \sqrt5$ in dashed lines, in $d=4$. The case
    $Q = 0$ is the Gibbons\hskip0.02cm-\hskip-0.02cm{}Hawking black
    hole, there is only the $r_{+1}(T,Q)$ solution, which is clearly
    unstable.  It is also plotted, in a gray line,
    the critical Davies radius as a
    function of $T$, ${r_+}_{\rm D}=\frac{1}{6\pi T}$.
    \label{fig:rtinfd4}}
    \end{figure}

To find the solutions of the canonical ensemble, we invert 
Eq.~\eqref{eq:Hawkbeta0d=4}
to yield $\left( \frac{1}{4\pi T}\right)(r_+^2- Q^2) 
- r_+^3= 0$, which is Eq.~\eqref{eq:rpinT}
for $d=4$.
This equation
can be solved analytically as it is a cubic equation, 
still, we do not present here the expression since it is cumbersome. 
Alternatively, the solutions can be 
analyzed qualitatively or solved numerically. 
One finds that the function $t_{\rm H}(r_+,Q)$
in Eq.~\eqref{eq:Hawkbeta0d=4}
has a maximum at 
$r_{+s}=\sqrt3\,Q$,
which is a saddle or critical point of the 
action of the black hole
and which we write as 
\begin{align}
{r_+}_{\rm D}=\sqrt3\,Q\,,
\label{r+daviesd=4}
\end{align} 
as in $d=4$ it gives the Davies
black hole horizon radius.
This saddle point of the 
action is at the temperature
given by
\begin{align}
T_{\rm D}=
\frac{1}
{6\sqrt{3}\,\pi Q}\,,
\label{Tdavies4d}
\end{align} 
see Eq.~\eqref{Tdavies}, when $d=4$.
From Eqs.~\eqref{r+daviesd=4} and
\eqref{Tdavies4d}, we can eliminate
$Q$ to give for a given $T$,
${r_+}_{\rm D}=\frac{1}{6\pi T}$,
or inverting, for a given $r_+$,
$T_{\rm D}=\frac{1}{6\pi r_+}$.
The temperature given in Eq.~\eqref{Tdavies4d}
is the Davies temperature, and it
is a result that can be extracted from \cite{Davies:1977,hut1977}.
One finds that for temperatures $T
\leq T_{\rm D}$ there are two black holes,
the solution
$r_{+1}(T,Q)$ and the solution $r_{+2}(T,Q)$, while
for $T > T_{\rm D}$
there are no black hole solutions.
The solution $r_{+1}(T,Q)$ is bounded in the interval $ r_{+e}
<r_{+1}(T,Q)\leq {r_+}_{\rm D}$, where $r_{+e}=r_{+1}(T\rightarrow
0,Q) = Q$ is the radius of the extremal black hole and ${r_+}_{\rm
D}=r_{+1}(T_{\rm D},Q) = \sqrt3\, Q $, so we can summarize for the
solution 1
\begin{equation}
\begin{aligned}
&r_{+1}=r_{+1}(T,Q),\quad\quad 0 <T \leq T_{\rm D},
\\
& q_1=Q,\quad\quad\quad\quad\quad\quad\,
r_{+e} <r_{+1}(T,Q)\leq {r_+}_{\rm D}.
\end{aligned}
\label{solutionr+1d=4}
\end{equation}
This solution,
$r_{+1}(T,Q)$, is an increasing monotonic
function of $T$.
The solution $r_{+2}(T,Q)$ is
in the interval
${r_+}_{\rm D}< r_{+2}(T,Q)<\infty$,
so we can summarize for the
solution 2
\begin{equation}
\begin{aligned}
&r_{+2}=r_{+2}(T,Q),\quad\quad 0 <T \leq T_{\rm D},
\\
& q_2=Q,\quad\quad\quad\quad\quad\quad\,
{r_+}_{\rm D}< r_{+2}(T,Q)<\infty,
\end{aligned}
\label{solutionr+2d=4}
\end{equation}
where, 
at $T \rightarrow 0$, the 
solution tends to infinity there. This solution,
$r_{+2}(T,Q)$, is a decreasing monotonic
function of $T$.
When the ensemble is only characterized
by the temperature $T$, with $Q$ vanishing, $Q=0$,
only the black hole 
$r_{+2}$ survives which is the
Gibbons\hskip0.02cm-\hskip-0.02cm{}Hawking black hole 
solution.
For $T>T_{\rm D}$ there are no black hole solutions and one
is left with hot flat space with electric charge
$Q$ dispersed at infinity, i.e.,
\begin{equation}
\begin{aligned}
&{\rm charged\; hot\;flat\;space},\quad\quad T_{\rm D}<T<\infty,
\\
& Q \;{\rm dispersed \;at}\; r=\infty,\quad\quad
0\leq r<\infty.
\end{aligned}
\label{solutionhot4d}
\end{equation}
The two solutions $r_{+1}(T,Q)$ and $r_{+2}(T,Q)$
are plotted as functions of the temperature
in Fig.~\ref{fig:rtinfd4}
 for two 
different values of the electric charge, with the features 
just mentioned. For $T>T_{\rm D}$ there are
no solutions, only hot flat space with electric charge
$Q$ at infinity.

The zero loop action 
of the canonical ensemble characterized by the
temperature $T$ and the electric charge $Q$
for $d=4$ can be
found using directly Eq.~\eqref{eq:actioninfinity},
i.e.,
\begin{align}
I_0[T,Q] = \frac{1}{2T}
\left(r_+
+ \frac{Q^2}{r_+}\right)
-\pi r_+^2
\,,
\label{eq:actioninfinityd=4}
\end{align}
where we have used $\mu=1$ and $\Omega_2=4\pi$.  The black hole
horizon radii $r_+$ that enter into this action are the $r_{+1}$ given
in Eq.~\eqref{solutionr+1d=4} or the $r_{+2}$ given in
Eq.~\eqref{solutionr+2d=4}.

\subsection{Thermodynamics in $d=4$}

With the solutions and the action of the canonical ensemble found, 
one can obtain the thermodynamics 
through the correspondence $F = T I_0$, where $F$ again is the 
Helmholtz free energy of the system.
From Eq.~\eqref{eq:actioninfinityd=4}, $F$ in $d=4$ is
$F =
 \frac12
\left(r_+
+ \frac{Q^2}{r_+}\right)
-T\,\pi r_+^2$,
which upon using Eq.~\eqref{eq:Hawkbeta0d=4}
gives
\begin{align}
F(T,Q)
=
\frac{1}{4}
\left(
r_+ 
+  \frac{3Q^2}{r_{+}}
\right),
\label{eq:Fd=4}
\end{align}
where $r_+$ should be envisaged
as  $r_+=r_+(T,Q)$, since it is one of
the solutions 
$r_{+1}(T,Q)$ or
$r_{+2}(T,Q)$, given in
Eq.~\eqref{solutionr+1d=4} or 
Eq.~\eqref{solutionr+2d=4},
respectively.
From the derivatives of the free energy, 
we obtain the entropy as 
$S = \pi r_+^2$, i.e., $S = \frac14 A_+$,
the electric potential, $\phi = \frac{Q}{r_+}$,
and 
the thermodynamic energy,
$E = \frac12
\left(r_+
+ \frac{Q^2}{r_+}\right)$, where we have used that $E = F + TS$.
These expressions are valid
for both solutions, $r_{+1}$ and $r_{+2}$.
The expression for the energy is precisely the expression for
the space mass $m$, so
$E = m$. The free energy of
Eq.~\eqref{eq:Fd=4} is then $F=m-TS$.

Here, the Smarr formula in $d=4$ is clearly
\begin{align}
m = \frac12\, TS + \phi Q\,,
\end{align}
see Eq.~\eqref{smarrd} for $d=4$.
Also, one has that the law
$dm =  TdS + \phi dQ$ holds, which ties 
the first law of black hole mechanics with the 
first law of thermodynamics.
The first law of black hole mechanics
is the expression from which Davies 
\cite{Davies:1977} started his thermodynamic analysis,
see also \cite{hut1977}.
We have started our analysis from
the canonical ensemble theory and the
Euclidean path integral approach with
the action of Eq.~\eqref{eq:actioninfinityd=4}, which yields 
naturally the first law of thermodynamics.

The heat capacity $C_Q$ of Eq.~\eqref{heatcapacityraw},
is for $d=4$ given by
\begin{align}
C_Q =
\frac
{2\pi r_+^2 \left(1-\frac{Q^2}{r_+^2}\right)} 
{3\frac{Q^2}{r_+^2}-1}=
\frac{m S^3  T}{\frac{\pi Q^4}{4} -
T^2
S^3}\,,
\label{heatcapacityraw4d}
\end{align}
where in the second equality we wrote the heat capacity in terms of
the thermodynamic variables $m$, $S$, $T$, and $Q$.
Note that $C_Q$ is a function of $T$ and $Q$, which are the
parameters that are controlled.
Thermodynamic stability is governed by the positivity of the heat
capacity, $C_Q\geq0$. From Eq.~\eqref{heatcapacityraw4d}, one finds
that the range of stability is $r_{+e}\leq r_{+} < {r_+}_{\rm D}$,
where $r_{+e}$ is the radius 
of the extremal black hole given by $r_{+e}=Q$
and 
${r_+}_{\rm D}$ is the Davies horizon radius 
given in Eq.~\eqref{r+daviesd=4}.
This range for $r_+$
corresponds to the solution $r_{+1}$, and so one has
\begin{align}
{\rm stability \,\, if}\; C_Q\geq0,\quad{\rm i.e.,}\quad
r_+=r_{+1}\,.
\label{stabilityRinfinityd=4}
\end{align}
Since  ${r_+}_{\rm D}=\sqrt3\,Q$, 
Eq.~\eqref{stabilityRinfinityd=4} is equivalent to
$Q\geq\frac1{\sqrt3}r_+$, i.e., one has
$\frac1{\sqrt3}r_+\leq Q\leq r_+$,
the latter term being the extremal case.
Now, the relation between the horizon radius, the
mass, and the electric charge of
the black hole is 
$r_+=m+\sqrt{m^2-Q^2}$, so $Q\geq\frac1{\sqrt3}r_+$
is the same as $Q \leq m\leq\frac{2}{\sqrt3} Q$,
which is another manner of writing the
condition for stability, and is 
the expression that can be found in \cite{Davies:1977},
see also \cite{hut1977}. The heat capacity goes
to zero at the extremal case 
$\frac{Q}{r_+}=1$.
Moreover, from Eq.~\eqref{heatcapacityraw4d}, one finds
that the range of instability is
${r_+}_{\rm D} <r_{+} <\infty$.
This range for $r_+$
corresponds to the solution $r_{+2}$, hence there is
\begin{align}
\hskip -0.3cm
{\rm instability \,\,if}\; C_Q<0,\quad{\rm i.e.,} \quad
r_+=r_{+2}\,.
\label{instabilityRinfinityd=4}
\end{align}
The inequality on the horizon radius for the 
case of instability can be rewritten as $0\leq Q< \frac1{\sqrt3}r_+$.
Note that
when the electric charge is zero, the heat capacity is negative
for all $r_+$,
indeed for $Q=0$ the heat capacity is 
$\frac{C}{r_+^2}
= -2 \pi$.
Note that $C_Q$ given in 
the second part of Eq.~\eqref{heatcapacityraw4d}
is the same formula found in \cite{Davies:1977} by 
performing in Eq.~\eqref{heatcapacityraw4d} the redefinitions 
$S \rightarrow 8\pi S$,
$T \rightarrow \frac{1}{8\pi}T$ and 
$\frac{C_Q}{8\pi} \rightarrow C_Q$,
in  \cite{hut1977} the conventions are yet different
from ours and from \cite{Davies:1977}.

\begin{figure}[h]
\centering
\includegraphics[width=\linewidth]{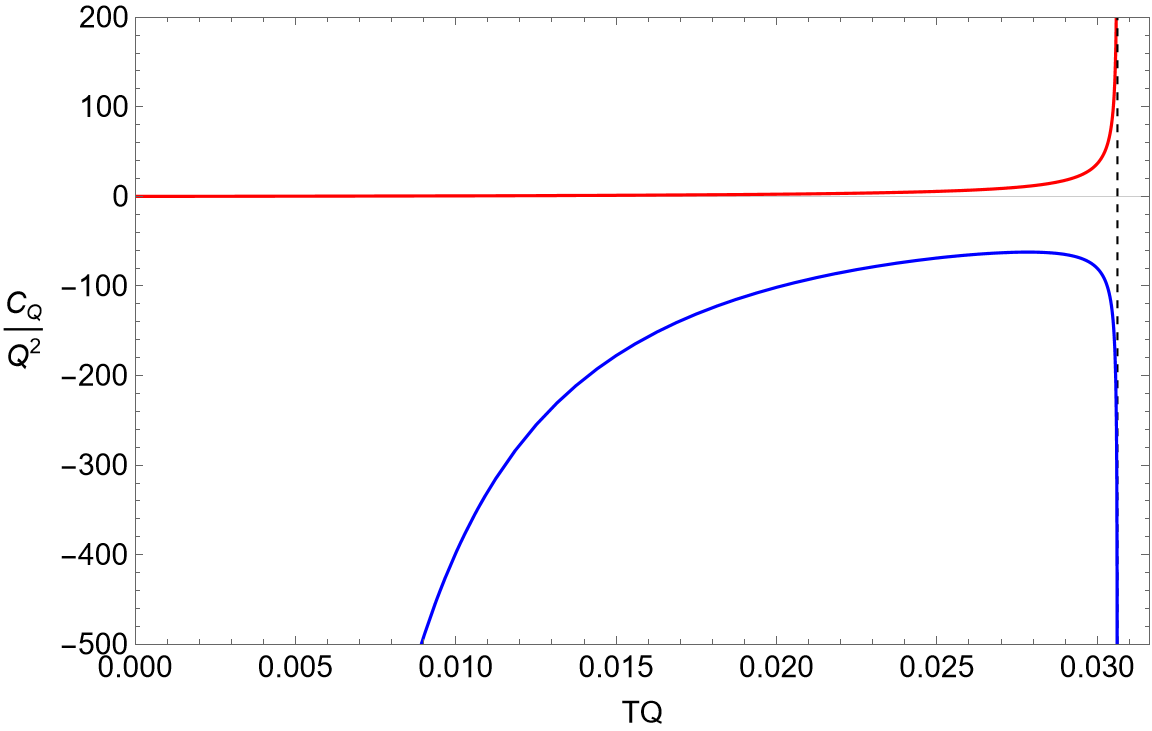}
\caption{The heat capacity $C_Q$ in $Q^2$ units,
$\frac{C_Q}{Q^2}$, is given as a function of the
temperature parameter $TQ$ in $d=4$, for the stable solution 
$r_{+1}$ in red and unstable solution $r_{+2}$ in blue. 
The heat capacity diverges for both solutions 
at the turning point $T_{\rm D} Q = \frac{1}{6\pi\sqrt{3}} = 0.03$, 
the latter equality being approximate.}
\label{CAq4dRinfty}
\end{figure}

The heat 
capacity $C_Q$
in units of $Q^2$, i.e.,
$\frac{C_Q}{Q^2}$, as a function
of the temperature
parameter, i.e.,  $T Q$
is plotted in Fig.~\ref{CAq4dRinfty}.
For each $T Q$, the heat capacity is double-valued, being 
positive for $r_{+1}$ in the red curve and negative for $r_{+2}$ in the 
blue curve. Therefore, the solution $r_{+1}$ is stable as it is 
expected from the increasing monotonic behavior of 
$r_{+1}$ with increasing temperature, while the solution 
$r_{+2}$ is unstable, having the opposite monotonic behavior. 
When $Q=0$, there is only the
$r_{+2}$ solution corresponding to the unstable 
Gibbons\hskip0.02cm-\hskip-0.02cm{}Hawking 
black hole solution. At the Davies point, corresponding to
$T_{\rm D} Q = \frac{1}{6\pi\sqrt{3}}$, 
the heat capacity goes to plus
infinity, for the solution $r_{+1}$ and to
minus
infinity for the solution $r_{+2}$.
If for some $T$, the configuration of the ensemble happens to be in 
the unstable $r_{+1}$ solution then it will transition to
the stable $r_{+2}$, since any thermal perturbations make the 
solution $r_{+2}$ run away from equilibrium. 
This happens for all temperatures
up to $T_{\rm D}$, where the two solutions coincide,
and for higher $T$, there are no more black hole
solutions. Thus, the point with temperature $T_{\rm D}$ 
characterizes a turning
point. It was stated by Davies that such point might be 
classified as a second order phase transition. However, this 
cannot be the case for the canonical ensemble, as discussed above, 
because only the stable solutions must be considered and the temperature 
$T_{\rm D}$ signals the upper limit of existence of the stable solution. 
Another way of looking at the Davies point, through the ranges of the 
horizon radius, is that it provides the relative scale
between $r_+$ and $Q$ at which 
one has black hole stability or metastability
in the canonical ensemble with a heat reservoir at infinity.
In \cite{Davies:1977}, Davies has plotted ${C_Q}\times Q$ in some units
of ${C_Q}$ and of $Q$ at constant mass $m$,
whereas here we have plotted $\frac{C_Q}{Q^2}\times TQ$,
where $TQ$ is a temperature parameter, as $Q$ is kept
constant in the calculation of ${C_Q}$.

The analysis of the favorable thermodynamic states
for $d=4$ does not differ from the analysis for generic
$d$ given above. We recapitulate the analysis here for
completeness.
There are two possible phases, the stable black hole $r_{+1}$
phase and the phase with hot flat space with electric charged 
at infinity.
For $T\leq T_{\rm D}$, both $r_{+1}$ and hot flat space
with electric charge at infinity can exist.
Since the free energy for hot flat space is zero
and the free energy for the black hole 
$r_{+1}$ is positive for all $T$ and $Q$,
hot flat space with electric charge at
infinity is the favored phase for $T\leq T_{\rm D}$.
In this range of temperatures, 
if 
the system is in the black hole phase, it will
settle upon perturbation in 
a hot flat space with charge at infinity phase
which has lower free energy, in the same way that supercooled
water phase changes into ice.
For $T>T_{\rm D}$, there are no black holes, 
hot flat space with electric charge at
infinity is the only phase.

An interpretation of the results in $d=4$ in terms of wavelengths
$\lambda$ of packets of thermal energy inside the cavity of a heat
reservoir at infinity also follows the analysis for generic $d$ given
above. Note that this interpretation goes beyond the formal
results we have found, since we have worked at the zero loop
approximation which does not treat quamtum matter
fields. Nevertheless, it is beneficial to give an interpretation.
The essential idea is that at a given $T$ and so at a given $\lambda$,
the small black hole is smaller than $\lambda$ and the radiation is
trapped outside, while the large black hole is larger than $\lambda$
and the radiation can escape the black hole.  For sufficient high $T$,
there is too much agitation in packets of energy with small wavelength
$\lambda$, and these packets wonder undisturbed by gravity in hot flat
space with the electric charge being deposited uniformly at infinity.
In Fig.~\ref{fig:rtinfd4}, the curve ${r_+}_{\rm D}=\frac{1}{6\pi T}$
is drawn in gray, but this is the definition of $\lambda=
\frac{1}{6\pi T}$ for $d=4$. And so, Fig.~\ref{fig:rtinfd4} describes
precisely the interpretation in terms of wavepackets given
above. Indeed, from small $T$ up to $T_{\rm D}$, the gray curve is
larger than the horizon radius of the smaller black hole, while it is
smaller, although of the same order, than the horizon radius of the
larger black hole. At $T_{\rm D}$, the gray curve and both solutions
meet. For larger temperatures than $T_{\rm D}$, there are no black
hole solutions.

A further comment is in order.  The first law of black hole mechanics
is the expression from which Davies \cite{Davies:1977} started his
analysis, see also \cite{hut1977}.  We have started our analysis from
the statistical mechanics canonical ensemble theory using the
Euclidean path integral approach and the action of
Eq.~\eqref{eq:actioninfinityd=4} rather than starting from the first
law of black hole thermodynamics.  In the Reissner-Nordstr\"om black
hole case in the canonical ensemble, as opposed to the Schwarzschild
case, there is true thermodynamics, since there are instances where
the system is thermodynamically stable.  This thermodynamic stability
of black holes for a heat reservoir at constant $T$ and $Q$ contrasts
with the thermodynamic instability of all electrically charged black
holes in a heat reservoir at constant $T$ and constant electric
potential $\phi$, i.e., Reissner-Nordstr\"om black holes in the grand
canonical ensemble.  This latter case was the case analyzed in
\cite{gibbhawk:1977} using the Euclidean path integral approach for
the grand canonical ensemble, where this instability was noticed but
there was no attempt to cure the problem.  The appropriate setting
that gives a meaningful path integral and a corresponding
thermodynamics is within the electrically charged canonical ensemble
rather than the grand canonical one.

\section{The case $d=5$: A typical higher-dimensional case
\label{sec:Daviesd5}}

\subsection{Solutions and action in $d=5$}

We present here the case with dimension $d=5$, as it is a
typical higher dimension, and it is the 
first possible extension of the results provided by Davies.

One must 
start from the canonical ensemble characterized by the
temperature $T$ and the electric charge $Q$
at infinity in $d=5$. 
The black hole solutions $r_+$
of the ensemble are taken from Eq.~\eqref{eq:Hawkbeta}
together with Eq.~\eqref{eq:betahawking}
which in $d=5$ give
\begin{align}
T=t_{\rm H}(r_+,Q),\quad\quad
t_{\rm H}(r_+,Q)=
\frac{r_+^2-\frac{4 Q^2}{3\pi r_+^2}}{2\pi r_+^3},
\label{eq:Hawkbeta0d=5}
\end{align}
where
$T$ is the temperature kept by the reservoir
at infinity and 
$t_{\rm H}(r_+,Q)$ is
the Hawking function in $d=5$.
When the electric charge of the reservoir at infinity
is zero, $Q=0$, then 
$t_{\rm H}(r_+,Q)=\frac{1}
{2\pi r_+}$, which is the Hawking temperature of a
Schwarzschild black hole in $d=5$.
The electric charge $Q$ is 
the electric charge kept by the reservoir
at infinity, and the black hole electric charge
$q$ 
must match it to have a consistent solution,
$q=Q$.

To find the solutions of the canonical ensemble, we invert 
Eq.~\eqref{eq:Hawkbeta0d=5}
to yield
$\left( \frac{1}{2\pi T}\right)(r_+^4-\frac{4}{3\pi} Q^2) 
- r_+^5 = 0$, which is Eq.~\eqref{eq:rpinT}
for $d=5$, a quintic
equation not easily  solvable analytically. 
 However, it can be 
analyzed qualitatively or solved numerically.
One finds that the function $t_{\rm H}(r_+,Q)$
in Eq.~\eqref{eq:Hawkbeta0d=4}
has a maximum at 
\begin{align}
    r_{+s}= \left(\sqrt{\frac{20}{3\pi}}\,Q\right)^\frac{1}{2}\,.
    \label{r+sdaviesd=5}
\end{align} 
which is a saddle point of the 
action of the black hole,
with a corresponding temperature at the reservoir
given by
\begin{align}
T_s=
\frac{2}
{
5\pi
\left(\sqrt{\frac{20}{3\pi}}\,Q\right)^{\frac12}
}
\label{Tdaviesd=5}\,.
\end{align}
From Eqs.~\eqref{r+sdaviesd=5} and
\eqref{Tdaviesd=5}, we can eliminate
$Q$ to give for a given $T$,
$r_{+s}=\frac{2}{5\pi T}$,
or inverting, for a given $r_+$,
$T_s=\frac{2}{5\pi r_+}$.
One finds that for temperatures $T
\leq T_s$, there are two black hole solutions, 
the solution
$r_{+1}(T,Q)$ and the solution $r_{+2}(T,Q)$,
while for $T > T_s$
there are no black hole solutions.
The solution $r_{+1}(T,Q)$ is bounded in the interval $ r_{+e}
<r_{+1}(T,Q)\leq r_{+s}$, where $r_{+e}=r_{+1}(T\rightarrow
0,Q) = \left(\frac{2}{\sqrt{3\pi}}Q\right)^{\frac{1}{2}}$ 
is the radius of the extremal black hole and
$r_{+s}=r_{+1}(T_s,Q) =
\left(\sqrt{\frac{20}{3\pi}}Q\right)^\frac{1}{2}$,
so we can summarize 
solution 1
in the form
$r_{+1}=r_{+1}(T,Q)$,
$q_1=Q$, with
$0 <T \leq T_s$ and 
$r_{+e} <r_{+1}(T,Q)\leq r_{+s}$.
 This solution,
$r_{+1}(T,Q)$, is an increasing monotonic
function of $T$.
The solution $r_{+2}(T,Q)$ is
in the interval $ r_{+s}
<r_{+2}(T,Q)< \infty$,
so we can summarize
solution 2 in the form
$r_{+2}=r_{+2}(T,Q)$,
$q_2=Q$, with
$0 <T \leq T_s$
and
$r_{+s}< r_{+2}(T,Q)<\infty$,
where the 
solution tends to infinity at $T \rightarrow 0$. 
This solution, $r_{+2}(T,Q)$, is a decreasing monotonic
function of $T$.
When the ensemble is only characterized
by the temperature $T$, with $Q$ vanishing, $Q=0$,
only the black hole 
$r_{+2}$ survives which has the
Gibbons\hskip0.02cm-\hskip-0.02cm{}Hawking 
black hole solution 
properties.
For $T>T_s$, there are no black hole solutions and one
is left with hot flat space with electric charge
$Q$ dispersed at infinity, i.e.,
one has
charged
hot
flat
space for $T_s<T<\infty$ with $Q$
 dispersed at
$r=\infty$.

The zero loop action 
of the canonical ensemble, which is characterized by the
temperature $T$ and the electric charge $Q$
at infinity, for $d=5$ can be
found using directly Eq.~\eqref{eq:actioninfinity},
i.e.,
\begin{align}
I_0[T,Q] = \frac{1}{2T}
\left(\frac{3\pi r_+^2}{4} 
+ \frac{Q^2}{r_+^2}\right)
-\frac{\pi^2 r_+^3}{2}\,,
\label{eq:actioninfinityd=5}
\end{align}
where we have used $\mu=\frac{4}{3\pi}$
and $\Omega_3=2\pi^2$.  The black hole
horizon radii $r_+$ that enter into this action are $r_{+1}$
or  $r_{+2}$.

\subsection{Thermodynamics in $d=5$}

With the solutions and the action
of the canonical ensemble found, 
one can obtain the thermodynamics 
through the correspondence $F = T I_0$, that comes from the 
zero loop approximation of the path integral, where $F$ again is the 
Helmholtz free energy of the system.
From Eq.~\eqref{eq:actioninfinityd=5},
$F$ in $d=5$  is
$
F = \frac{1}{2}
\left(\frac{3\pi r_+^2}{4} 
+ \frac{Q^2}{r_+^2}\right)
-T\frac{\pi^2 r_+^3}{2}\,,
$.
Substituting $T$ for $t_{\rm H}$, see Eq.~\eqref{eq:Hawkbeta0d=5}, 
we obtain for
the free energy the expression
\begin{align}
F(T,Q)
=
\frac{\pi}{8}
\left(
r_+^2
+ \frac{20 Q^2}{3\pi r_+^2}
\right),
\label{eq:Fd=5}
\end{align}
where $r_+$ should be envisaged
as  $r_+=r_+(T,Q)$, since it is one of
the solutions 
$r_{+1}(T,Q)$ or
$r_{+2}(T,Q)$.
Thus, the Helmholtz free energy $F$
for each solution is a function only of $T$ and $Q$,
namely, 
$F(T,Q,r_{+1}(T,Q))$ and $F(T,Q,r_{+2}(T,Q))$.
By computing the derivatives of the free energy,
we can obtain the entropy
as $S = \frac14 A_+$,
$A_+ = 2\pi^2 r_+^3$,
the 
thermodynamic
electric potential, which is $\phi = \frac{Q}{r_+^2}$,
and
the energy, which is 
$E = \frac{3\pi r_+^2}{8} + \frac{Q^2}{2r_+^2}$, 
where we have used $E = F - TS$.
These expressions are valid
for both solutions, $r_{+1}$ and $r_{+2}$.
The energy has precisely the expression for
the space mass $m$, so
$E=m$.
The free energy of
Eq.~\eqref{eq:Fd=5} is then $F=m-TS$.

Here, in $d=5$ the Smarr formula takes the form
\begin{align}
m = \frac23\, TS + \phi Q\,.
\end{align}
Also, one has that the law
$dm =  TdS + \phi dQ$
holds. This is the first law of black hole mechanics, 
which is also the first law of black hole thermodynamics. 
And in fact, the first law of black hole thermodynamics is valid
in the electrically charged case
for the  instances where the
system is thermodynamically stable.

The heat capacity of Eq.~\eqref{heatcapacityraw}
is in $d=5$ given by
\begin{align}
C_Q =&
\frac
{3\pi^2  r_+^3 
\left(1-\frac{4}{3\pi}\frac{Q^2}{r_+^4}\right)} 
{2\left(\frac{20}{3\pi}\frac{Q^2}{r_+^4}-1\right)}\nonumber\\
 =&
\frac{m S^3  T}{\frac{7\pi^2}{36}Q^4
\left(\frac{2S}{\pi^2}\right)^{-\frac{1}{3}} 
+ \frac{\pi^4}{4^3}\left(\frac{2S}{\pi^2}\right)^{\frac{7}{3}} -
T^2
S^3}\,,
\label{heatcapacityraw5d}
\end{align}
where in the second equality we wrote the heat capacity in terms of
the thermodynamic variables $m$, $S$, $T$, and $Q$.
The heat capacity must be seen as a function of $T$ and $Q$, with
$r_+$ being given by either the solutions $r_{+1}(T,Q)$ and $r_{+2}(T,Q)$,
or as well $m=m(T,Q)$ and $S=S(T,Q)$.
In order to have thermodynamic stability, the heat capacity must
be positive, i.e., $C_Q \geq0$, which is accomplished by the range
$r_{+e}\leq r_+\leq r_{+s}$ or in terms of electric charge
$\left(\frac{3\pi}{20}\right)^{\frac12} r_{+}^2\leq
Q\leq\left(\frac{3\pi}{4}\right)^{\frac12} r_{+}^2$, with $r_{+s}$
given in Eq.~\eqref{r+sdaviesd=5}. This range is precisely the one of
the solution $r_{+1}$, and so the solution $r_{+1}$ is stable.  For
the remaining range, satisfied by the solution $r_{+2}$, the heat
capacity is negative, thus the solution $r_{+2}$ is unstable.
The heat capacity, when the electric charge is zero, is negative for
all $r_+$,
given by
$\frac{C}{r_+^3} = -\frac{3\pi^2}{2}$.
The heat capacity has the feature that diverges for each solution
at $T_s$, which is a turning point of the two solutions.
The heat capacity goes to zero at the extremal case
$\frac{Q}{r_+^2}=\left(\frac{3\pi}{4}\right)^{\frac12}$.  One can
also
infer that the solution is stable if the radius $r_+$
increases as the temperature increases,
yielding the same analysis above.

The analysis of the favorable thermodynamic states for $d=5$ follows
the same reasoning as for generic $d$.  There is the small stable
black hole phase and the hot flat space with electric charge at
infinity phase. Depending on the temperature, either the latter is favored or
it is the only phase.

An interpretation of the results in $d=5$ in terms of wavelengths
$\lambda$ of packets of thermal energy inside the cavity of a heat
reservoir at infinity also follows the analysis for generic $d$ given
above. We shall not repeat it here.

\section{Conclusions}
\label{sec:Concl}

We have shown that the Gibbons\hskip0.02cm-\hskip-0.02cm{}Hawking
Euclidean path integral approach for electrically charged black holes
in the canonical ensemble has in its core the Davies'
thermodynamic theory of black holes.
By framing these results within the rigorous structure of statistical
mechanics and ensemble theory which provide a deeper description of
the physics world, our work places Davies' theory and its
generalizations on a firm theoretical footing, and also offers new
insights into the thermodynamic stability of black holes.

To determine this connection,
we have computed the canonical partition function in the
Gibbons\hskip0.02cm-\hskip-0.02cm{}Hawking
Euclidean path integral approach for a Reissner-Nordstr\"om black hole
in $d$ dimensions.  The Euclidean action that enters into the path
integral consists of the Einstein-Hilbert-Maxwell action with the
Gibbons\hskip0.02cm-\hskip-0.02cm{}Hawking-York boundary term and an
additional Maxwell boundary term so that the canonical ensemble is
well-defined.  We have assumed that the heat reservoir resides at the
boundary of space, at infinite radius, where the temperature $T$ is
fixed as the inverse of the Euclidean proper time length at the
boundary, and also the electric charge is fixed by fixing the electric
flux at the boundary.
The zero loop approximation was then performed by giving the
expressions for the metric and the Maxwell tensor of the
Einstein-Maxwell system, obtaining the black hole solutions of the
ensemble, $r_+(T,Q)$.  We showed that there are two solutions for
temperatures below a critical value.  The smaller black hole solution 
is stable, while the larger one is unstable.  The two solutions meet at a
saddle or critical point, given formally by $r_{+s}=r_+(T_s,Q_s)$.
Above the saddle value for the temperature, there are no black
hole solutions, only hot flat space with electric charge
dispersed at infinity.
The thermodynamics of the system follows, since the canonical
partition function connects directly to the Helmholtz free energy.
The entropy obtained from the free energy is the Bekenstein-Hawking
entropy, the electric potential is the usual Coulombic potential, and
the thermodynamic energy is the mass of the black hole. The
thermodynamic stability is controlled by the heat capacity at constant
electric charge, which must be positive for stable solutions and
negative for unstable solutions. There is a turning point
precisely at the saddle values $T_s$ and $Q_s$.  The solution
with smaller radius is thermodynamically stable while the solution with 
larger radius is thermodynamically unstable.  
The Smarr formula relating mass, temperature, entropy,
electric potential, and electric charge follows naturally.  In
addition, the first law of thermodynamics reduces to the first law of
black hole mechanics, which, strictly speaking, is valid
only for the case of the stable solution.
We have studied the favorable phases comparing the free energies of
the stable black hole phase and the hot flat space with electric
charge at infinity phase. We have obtained that hot flat space with electric
charge at infinity is favorable throughout the configuration
space. If, for some reason, the system finds itself in the black hole
phase, it will make a transition to
hot flat space with electric charge at infinity.
This fact is due to the 
black hole phase not being a global minimum
of the free energy $F$, the global miminum of $F$
being hot flat space with charge at infinity.
Since the free energy 
of these two phases never intersects, one
cannot call this a first order phase transition.
However, if one includes the matter sector, it may
be possible that a first order phase transition exists between black
hole and matter.
We have given an
interpretation for the solutions and their stability in terms of
wavelengths of energy packets.  By considering
the dimension $d=4$, we then showed
that Davies' thermodynamic theory of black holes follows directly from
the whole formalism presented.  Davies' starting point for the theory
was the first law of black hole thermodynamics, our starting point was
the path integral approach with its action, and from it, we deduced the
first law of black hole thermodynamics and the critical points found
by Davies.  We also applied the theory to the case $d=5$. Our analysis
generically points towards the equivalence between the black hole
mechanics and black hole thermodynamics through the canonical
ensemble with an appropriate heat reservoir at infinity.

\section*{Acknowledgements}
We thank financial support from Funda\c c\~ao para a Ci\^encia e
Tecnologia - FCT through the project~No.~UIDB/00099/2020 and
project~No.~UIDP/00099/2020. TF acknowledges a grant from FCT 
no. RD0970.

\end{document}